\documentclass{aa}
\usepackage{graphics,epsfig}

\newcommand{\kms}{km~s$^{-1}$}
\newcommand{\cm}{cm$^{-2}$}

\newcommand{\dla}{ damped Lyman-$\alpha$ }

\begin{document}
\title{ HI 21cm imaging of a nearby Damped Lyman-$\alpha$ system}
\titlerunning{HI 21cm imaging of a nearby Damped Ly-$\alpha$ System}
\author{Jayaram N. Chengalur\inst{1}\thanks{chengalu@ncra.tifr.res.in},
        Nissim Kanekar \inst{1}\thanks{nissim@ncra.tifr.res.in}
}
\authorrunning{Chengalur \& Kanekar}
\institute{National Centre for Radio Astrophysics, 
Post Bag 3, Ganeshkhind, Pune 411 007}
\date{Received mmddyy/ accepted mmddyy}
\offprints{Jayaram N. Chengalur}

\maketitle
\begin{abstract}
	We present Giant Metrewave Radio Telescope (GMRT) HI~21cm
emission images of the $z=0.009$ damped Lyman-$\alpha$ (DLA) absorber
towards the QSO~HS~1543+5921. The DLA has been earlier identified
as a low surface brightness galaxy SBS~1543+593~(\cite{reimers98,bowen01a})
with small ($\sim 400$~pc) impact parameter to the QSO line of sight.
The extremely low redshift of the absorber allows us to make, for the first 
time, spatially resolved images of the 21cm emission; besides the HI mass, 
this also enables us to determine the velocity field of the galaxy and, hence, 
to estimate its dynamical mass. 
	We obtain a total HI mass of $\sim 1.4 \times 10^9$ M$_\odot$, 
considerably smaller than the value of M$_{\rm HI}^*$ determined from 
blind 21cm emission surveys. This continues the trend of low HI mass in 
all low redshift DLAs for which HI emission observations have been attempted. 
We also find that the QSO lies behind a region of low local HI column density 
in the foreground galaxy. This is interesting in view of suggestions that 
DLA samples are biased against high HI column density systems. Finally, the 
dynamical mass of the galaxy is found to be M$_{\rm dyn} \sim 5 \times 10^9$~M$_\odot$.
\end{abstract}

\section{Introduction}
\label{intro}

	Neutral gas at high redshifts is easiest to detect in absorption
against bright background sources. Not surprisingly, most of what we know 
about the content and evolution of neutral gas in the universe comes from 
the study of absorption lines seen in the spectra of distant QSOs. 
The number density (per unit redshift) of these absorption lines is
a strong function of the column density, with low column density 
($N_{\rm HI} \sim 10^{13}$~atoms~cm$^{-2}$) systems being several orders
of magnitude more common than high column density ($N_{\rm HI} \ga 
10^{20}$~atoms~cm$^{-2}$) systems. Nonetheless, the bulk of the 
neutral gas at high redshift is contained in these rare high HI column 
density absorbers. It is principally for this reason that these objects (called 
\dla~systems or DLAs) are obvious candidates for the precursors of today's 
galaxies. Further, the gas mass in DLAs at $z \sim 3$ is comparable to the 
stellar mass in galaxies at $z=0$ (e.g. \cite{storrie96,storrie2000}), consistent 
with the idea that the absorbers have converted their gas into stars in the 
intervening period. Understanding the nature of DLAs at different redshifts 
is clearly important in tracing the evolution of galaxies; for this reason, 
the absorbers have been the subject of considerable study over the last two decades.

	Unfortunately, since QSOs are  point sources, optical absorption studies 
alone are unable to constrain the transverse size of the absorbers. The typical 
size and mass of DLAs have hence long been controversial issues. Traditionally,
 DLAs have been modelled as large proto-spirals (\cite{wolfe86}). Some support for 
this model comes from the shapes and widths of the absorption profiles produced 
by ions such as SiII (which are associated with neutral HI). The large velocity 
widths ($\Delta V \sim 300$~km~s$^{-1}$) and the asymmetric shapes of these lines 
have been successfully modelled as arising from gas in a thick spinning disk 
(\cite{prochaska97}, 1998). However, models involving infall or random motions 
of smaller gas clouds have also been found to succesfully reproduce the observed 
velocity profiles (\cite{haehnelt98, mcdonald99}).

	At low redshifts, the galaxies in which the damped absorption arises 
can be directly studied by ground-based or HST observations. Contrary to 
expectations, low redshift DLAs appear to be associated with a wide variety 
of galaxy types, including dwarf and low surface brightness (LSB) galaxies (e.g. 
\cite{lebrun97,nestor01}) and not exclusively (or even predominantly) with spiral 
galaxies. Besides this, the majority of DLAs tend to have low metallicities 
($\sim 0.1$ solar) at all redshifts, with very little evolution in their metallicity
with redshift (\cite{pettini99}); further, they also do not show the expected 
$\alpha$/Fe enrichment pattern expected for spiral galaxies (\cite{centurion2000}), 
suggesting a different star formation history than that of spirals. Finally, 
21cm absorption studies (\cite{chengalur00,kanekar01a}) have shown that the majority 
of DLAs have high spin temperatures ($T_{\rm s} \sim 1000$~K), far higher than those 
observed in the Milky Way or local spirals ($T_{\rm s} \sim 200$~K). All these
results indicate that damped absorption is likely to originate in {\it all} 
types of galaxies and not merely in luminous disk systems.

	Of course, the above results do not preclude the possibility that the 
galaxies responsible for the damped absorption are indeed massive gas-rich disks, 
but have not undergone much star formation and hence have both low luminosities as 
well as low metallicities. Such low surface brightness systems could well have 
a larger fraction of the warm phase of neutral HI and hence, a high spin 
temperature (\cite{chengalur00}). However, for systems at low redshift, 21cm 
emission studies can be used to get direct estimates of the HI mass of the 
absorbers; one can thus directly test the above hypothesis, that DLAs are 
massive, low luminosity galaxies. Deep searches for HI emission from 
two nearby DLAs have resulted in non-detections (\cite{kanekar01b,lane00}). In 
both cases, the 3~$\sigma$ upper limit to the HI mass is $\sim$ 1/3 the HI mass 
of an L$_{*}$ spiral. Thus, in these two cases at least, the absorption does 
not arise in an optically faint, but extremely gas rich galaxy.

	In this paper, we discuss Giant Metrewave Radio Telescope (GMRT) 21cm 
emission observations of a third low redshift DLA, the $z=0.009$ absorber towards
the QSO~HS~1543+5921. The damped absorption has been identified as arising in
a low surface brightness galaxy SBS~1543+593 (\cite{reimers98, bowen01a}).
HI emission has also been detected from this galaxy using the Bonn Telescope 
(\cite{bowen01b}).  The extremely low redshift of the absorber allows us to make,
for the first time, spatially resolved images of the 21cm emission; besides the 
HI mass, this also enables us to determine the velocity field of the galaxy and, 
hence, to estimate its dynamical mass. 

	The GMRT observations are detailed in Sect.~\ref{sec:obs}, while the 
results are presented in Sect.~\ref{subsec:results} and discussed in  
Sect.~\ref{subsec:discuss}. Throughout the paper, we use a Hubble constant of 
H$_{\rm 0}$=~75~km~s$^{-1}$~Mpc$^{-1}$, i.e. a distance to SBS~1543+593 of 38~Mpc.

\section{Observations}
\label{sec:obs}

	SBS~1543+593 was observed on 31~December~2000 using the GMRT.
The observing setup gave a total of 128 spectral channels over the 
total bandwidth of 2~MHz, corresponding to a channel separation of 
$\sim 3.1$~\kms. The standard calibrators 3C48 and 3C286 were observed 
at the start and end of the observing run and used to calibrate 
the visibility amplitudes and the bandpass shape. Phase calibration 
was done using observations of the nearby continuum sources 1438+621 
and 1549+506, one of which was observed once every 35 minutes.

	The data were converted to FITS from the raw telescope format
and analysed in classic AIPS. The brightest continuum source in the 
field of view is the steep spectrum source 7C~1543+5920. The flux as 
measured at GMRT is 42~mJy, in excellent agreement with that obtained in 
the NVSS (\cite{condon98}). The source is at a distance of 2.2$^{'}$ 
(corresponding to an impact parameter of $\sim 24.1$~kpc) from SBS~1543+593. 
No statistically significant 21cm absorption is seen against this source; 
the 3~$\sigma$ limit to the optical depth after smoothing to a velocity 
resolution of 10~\kms~is 
$\sim 0.1$. This yields column density limits of $\sim 1.8\times 
10^{20}$~\cm~and $\sim 1.8\times 10^{21}$~\cm~for HI with spin temperatures 
of 100~K and 1000~K respectively, along this line of sight. As an aside, we 
note that 7C~1543+5920 has
a spectral index of $\sim -0.5$ and is, at best, marginally resolved on our 
longest baselines -- the upper limit on its angular size is $\sim 1^{''}$.  
Given its relatively high galactic latitude ($\sim 46^{o}$), it is likely 
that  7C~1543+5920 is a compact steep spectrum (CSS) source.

	The GMRT has a hybrid configuration; 14 of its 30 antennas are 	
located in a central compact array with size $\sim 1$~km ($\sim 5$~k$\lambda$ 
at 21cm) while the remaining antennas are distributed in the three arms 
of a Y configuration, giving a maximum baseline of $\sim 25$~km ($\sim 
120$ k$\lambda$ at 21cm). The baselines obtained from antennas in the 
central square are similar in length to those of the VLA in its D configuration, 
while the arm antennas 
provide baselines similar in length to the B array of the VLA. A single 
observation with the GMRT hence yields information on both large and small 
angular scales\footnote{Note, however, that the VLA in its D array has a 
larger number of short spacings than the central compact array of the GMRT.}. 
The search for absorption against the compact source 7C~1543+5920 was made 
using all available baselines. However, the HI emission from SBS~1543+593 is 
heavily resolved on the longest baselines. To study this emission, cubes 
were hence made at a variety of UV ranges, viz. 0-5~k$\lambda$, 
0-10~k$\lambda$ and 0-20~k$\lambda$, corresponding to spatial resolutions
of 39$^{''} \times 39^{''}$, $29^{''} \times 25{''}$ and $13^{''} \times 11^{''}$.
The two low resolution cubes were CLEANed, using the AIPS task 
IMAGR. At the highest resolution, the signal to noise ratio is
too low and the emission too diffuse for CLEAN to work reliably.
However, at the signal to noise ratio of the map at this resolution, 
convolution by the dirty beam does not greatly degrade the dynamic range 
(or image fidelity). The morphology should hence be accurately traced,
apart from an uncertainty in the scaling factor (which is discussed 
in more detail below).

\section{Results and Discussion}
\subsection{Results}
\label{subsec:results}

\begin{figure*}[t!]
\epsfig{file=uv5m0.eps,width=3.0in}
\vskip -3.6in
\hskip  3.25in \epsfig{file=uv10m0.eps,width=3.0in}

\hskip 0.12in  \epsfig{file=uv20m0.eps,width=3.0in}
\vskip -3.25in
\hskip  3.25in\epsfig{file=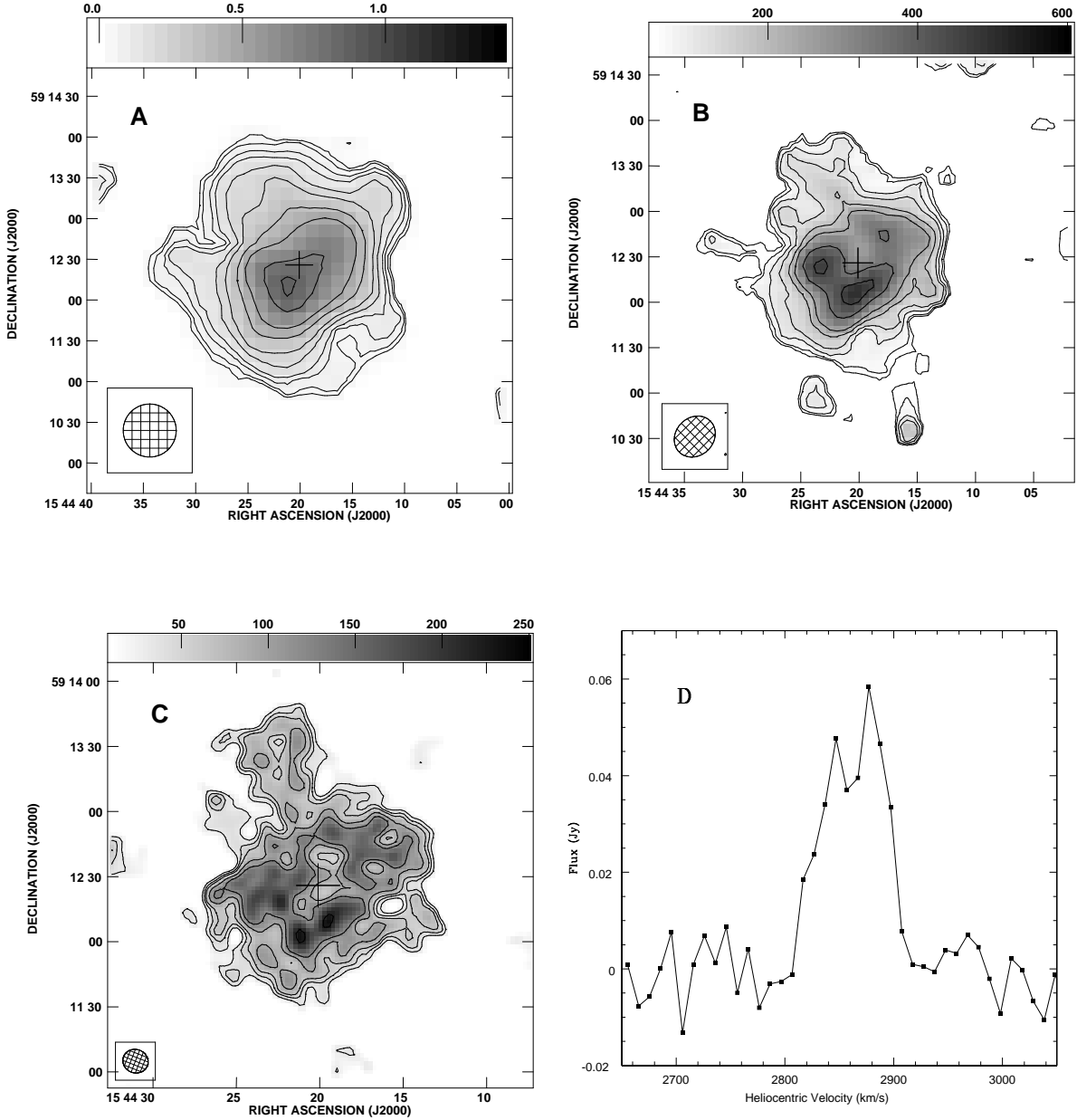,width=3.0in}
\caption{ Integrated HI emission maps of SBS1543+593. The QSO position
          is marked by a cross in all panels. See Sect.~\ref{sec:obs} for
          a discussion of the conversion to atoms cm$^{-2}$. 
          [A]~$39^{''} \times 39^{''}$ resolution. The contours are 0.03, 
              0.045, 0.06,0.09, 0.12, 0.18, 0.24, 0.36, 0.48, 0.72 and 
              0.85~Jy~Beam$^{-1}$~\kms.A $3\sigma$ feature, 2 channels wide
              would have an integrated flux of 0.07~Jy~Beam$^{-1}$~\kms
          [B]~$29^{''} \times 25^{''}$ resolution. The contours are 0.045,
              0.06, 0.09, 0.12, 0.18, 0.24, 0.36 and 0.45~Jy~Beam$^{-1}$~\kms.
              A $3\sigma$ feature, 2 channels wide would have an integrated
              flux of 0.06~Jy~Beam$^{-1}$~\kms
          [C]~$12^{''} \times 11^{''}$ resolution. The contours are 0.03,
              0.045, 0.06, 0.09, 0.12 and 0.22~Jy~Beam$^{-1}$~\kms.
              A $3\sigma$ feature, 2 channels wide would have an integrated
              flux of 0.045~Jy~Beam$^{-1}$~\kms
	  [D]~Integrated HI 21cm emission profile for SBS~1543+593. The
              profile is derived from the GMRT observations, using the 
	      $29^{''} \times 25^{''}$ resolution data. The spectrum has
              been smoothed to a resolution of $\sim$ 10~\kms. }
\label{fig:mom0}
\end{figure*}

	Fig.~\ref{fig:mom0} shows maps of the integrated HI emission
(obtained using the AIPS task MOMNT) at the three different resolutions
discussed above.  The QSO position is marked by a cross in all 
three maps. The centre of the optical galaxy is very close to the QSO
position; the QSO is located only $2.4^{''}$~NNE of the galaxy 
centre (\cite{reimers98}). The low resolution map shows that the HI 
distribution is asymmetric with the peak emission displaced to the 
south-east of the galaxy centre. The higher resolution maps show 
clearly that the HI distribution is concentrated in a ring-like 
structure, with HI emission actually being depressed at the centre of the 
galaxy. The HI ring is coincident with the faint spiral arms seen 
in the optical images (\cite{reimers98, bowen01a}). Many of the 
irregular patches of emission seen in the HST image (\cite{bowen01a}) 
also appear to be associated with peaks of HI emission. The HI 
concentration at $15^h44^m20^s \, , \, 59^d12^{'}09^{''}$ corresponds 
to the HII region whose spectrum is given in Reimers \& Hagen (1998). The 
velocity we measure  at the location of this HII region is $2855\pm 6$~\kms. 
In addition to the inner HI ring, there are also spurs in the  HI emission 
(see Figs.~\ref{fig:mom0}[B] \& [C]) towards the north and south; these may mark 
the beginning of faint outer spiral arms. 

\begin{figure*}[t!]
\epsfig{file=uv10m1.eps,width=3.0in}
\vskip -3.32in
\hskip  3.25in \epsfig{file=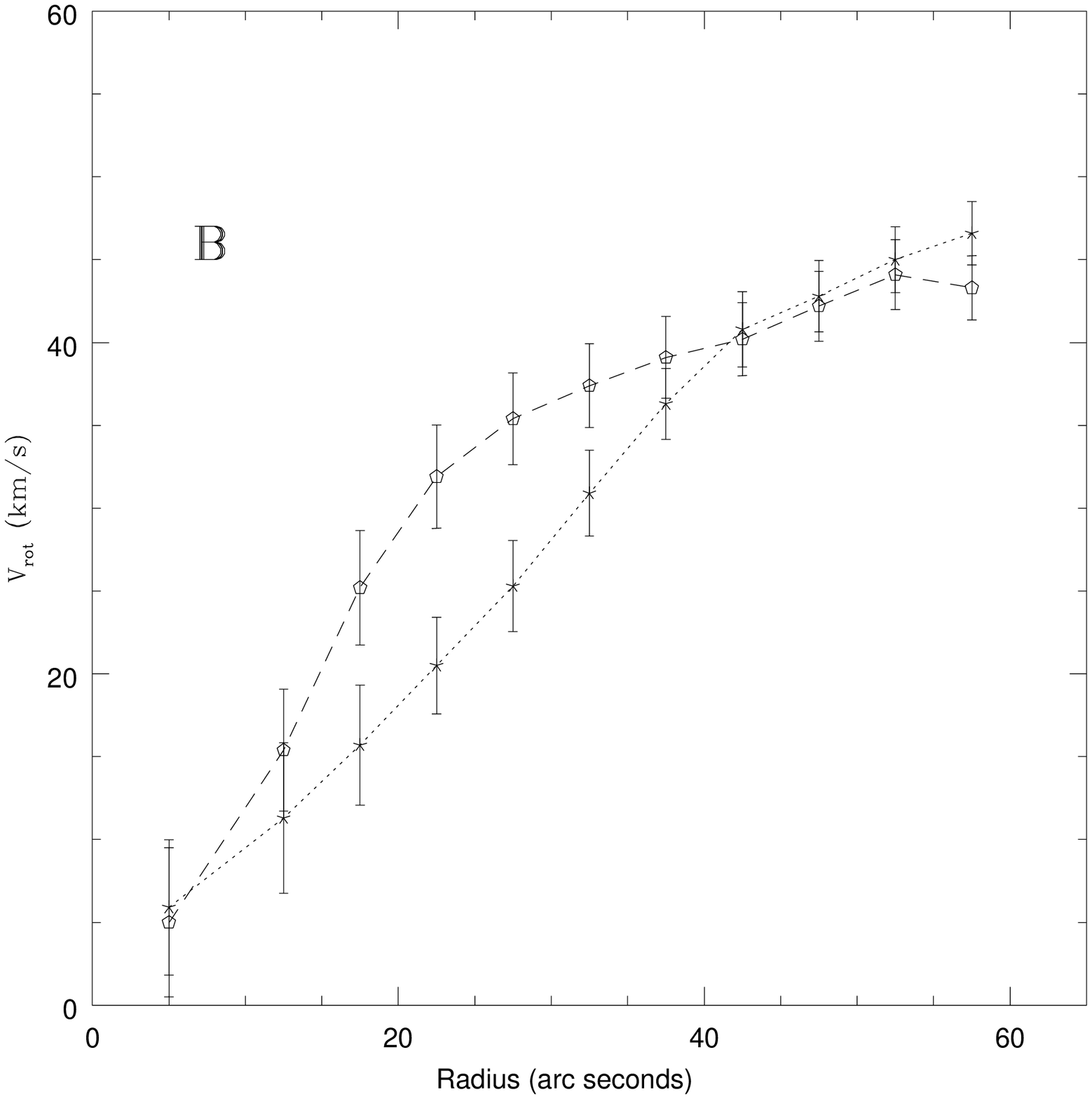,width=3.0in}
\caption{ [A]~The velocity field of SBS~1543+593 derived from the
              $29^{''} \times 25^{''}$ resolution cube. The velocity
              contours go from 2830~\kms~to 2900~\kms~and are
              spaced 5~\kms~apart. Note that the contours are straight
              on the approaching (lower velocity) side and curved on
              the receding (higher velocity) side. The QSO position
              (which is only 2.25$^{''}$ from the optical centre of
              the galaxy) is marked with a cross.
          [B]~The rotation curve derived from the velocity field shown
              in~[A]. The curve for the approaching side is marked by
              stars, and for the receding side by open pentagons.
              The rotation curve for the approaching side can be seen to rise
              approximately linearly, while that for the receding side 
	      flattens out.
         }
\label{fig:mom1}
\end{figure*}

 	Measurement of the integrated flux corresponding to weak extended 
line emission can be a non-trivial problem (see, for example, \cite{jorsater95,
rupen99}). However, as a rule of thumb, (apart from serious zero spacing
problems), deeply cleaned images give a fairly reliable estimate of the
total flux. The integrated fluxes that we get from the cleaned
$39^{''}\times39^{''}$ and $29^{''}\times25^{''}$ data cubes are 
$4.0\pm 0.4$~Jy~\kms~and $3.6\pm 0.4$~Jy~\kms. These are in excellent agreement 
with the single dish measurement of $4.0\pm 0.4$~Jy~\kms~(\cite{bowen01b}).
From the area of the clean beam in these two low resolution images, the 
conversion from 1~Jy~Beam$^{-1}$~\kms~to atoms~cm$^{-2}$ is
$7.3\times 10^{20}$ atoms~cm$^{-2}$ and $1.5\times 10^{21}$~atoms~cm$^{-2}$ 
respectively. Since the highest resolution map (Fig.~\ref{fig:mom0}[C]) is uncleaned, 
there is no obvious way to convert the units  from Jy~Beam$^{-1}$ to Jy. If one simply 
uses the area of the best fit Gaussian to the main lobe of the dirty beam to 
convert from Jy~Beam$^{-1}$ to Jy, one gets a total flux of $\sim 4.4$~Jy~\kms. This 
must be an overestimate, since it is clear from Fig.~\ref{fig:mom0}[C] that a 
considerable fraction of the smooth emission seen in Fig.~\ref{fig:mom0}[A] has been 
resolved out but the estimated flux in the higher resolution image is {\it higher} 
than that in the lower resolution ones. Given this problem in scaling for 
the highest resolution image, we estimate the column density at the QSO
position only from the two lower resolution maps. The column densities are 
$5.9\times 10^{20}$~atoms~cm$^{-2}$ and  $4.9 \times 10^{20}$ atoms~cm$^{-2}$ 
for the $39^{''} \times39^{''}$ and the $29^{''}\times25^{''}$ resolutions,
respectively. The integrated emission profile of SBS~1543+593, as derived 
from the $29^{''} \times 25^{''}$ resolution data (smoothed to a velocity 
resolution of 10~\kms), is shown in  Fig.~\ref{fig:mom0}[D]. The systemic 
velocity as measured from the profile is 2862$\pm$~10~\kms. This agrees
(within the errors of the two measurements) with the velocity of
2868$\pm$~2~\kms measured from the Bonn spectrum by \cite{bowen01a}.

	The velocity field of SBS~1543+593 (derived using the AIPS task
MOMNT on the $29^{''}\times25^{''}$ resolution cube) is shown in 
Fig.~\ref{fig:mom1}[A]. The iso-velocity contours
are asymmetric; they are straight on the approaching side and curved 
on the receding side. This type of velocity field has been dubbed 
`kinematically lopsided' (\cite{swaters99}). As noted earlier, the
galaxy is also morphologically lopsided. The rotation curve was derived
separately for the approaching and receding sides using the AIPS task
GAL. During these fits, the galaxy centre was kept fixed at the optical
centre. The systemic velocity was kept fixed at the value of 2870~\kms,
the inclination at $50^o$ and the position angle (of the receding half
of the major axis, measured east of north) at $-16^o$ (all of which 
were obtained from an initial global fit to the velocity field). The 
inclination and the position angle are in good agreement with our 
estimates from the optical image presented in Reimers \& Hagen (1998).

	The derived rotation curves are shown in Fig.~\ref{fig:mom1}[B].
As expected from the iso-velocity contours, the rotation curve is
almost linear on the approaching side, while it tends to flatten out on 
the receding side. The maximum (inclination corrected) rotation speed 
in SBS~1543+593 is $\sim 45$~\kms. The rotation curve can be measured
out to a radius of $\sim 60^{''}$, corresponding to a linear distance 
of 11~kpc.  The implied dynamical mass is then 
M$_{\rm dyn} \sim 5 \times 10^{9}$~M$_\odot$.

\subsection{Discussion}
\label{subsec:discuss}

	The flux that we measure for SBS~1543+593 corresponds to a 
total HI mass of $1.4 \pm 0.14\times 10^9$M$_\odot$, in excellent agreement
with the single dish measurement of Bowen et al. (2001b). This is thus
the third DLA whose HI mass is significantly less than that of an L$_*$
 spiral. Besides these, the galaxy NGC~4203~(which lies in front of the QSO
Ton~1480, and is likely to be a DLA; \cite{miller99}) has also been found to
have a low HI mass. Thus, all low redshift DLAs (or candidate DLAs) for which 
HI emission observations have been attempted have masses less than that of an 
L$_{*}$ spiral. 

	The derived HI masses are also considerably less than the M$_{\rm HI}^{*}$ 
($\sim 7 \times 10^{9}$ ~M$_\odot$) obtained from Schecter function fits to 
the HI mass function determined in blind HI surveys (\cite{zwaan97, rosenberg01}). 
It also follows from these mass functions that large galaxies make the major 
contribution to the local HI mass density. However, in order to determine 
the contribution of galaxies of different HI masses to the cross-section for 
DLA absorption, one also needs to know the typical sizes of their HI disks. 
Rao \& Turnshek (1998) deduced, based on determinations of the typical sizes 
of HI disks made by Rao (1994), that bright spiral galaxies make the major 
contribution to the $z=0$ cross-section for DLA absorption. However, more 
recent determinations of the HI size and mass distribution of galaxies indicate 
that the cross-section for DLA absorption is, in fact, not dominated by large spiral 
galaxies (\cite{rosenberg01,zwaan01}), and that a sample of low redshift DLAs 
should contain a large variety of galaxy types. This is consistent with the 
optical and HI observations of low $z$ DLAs.

	It would be interesting to compare the velocity profiles of low ionization 
metal lines in the HS~1543+5921 spectrum with the large scale kinematics of 
SBS~1543+593. Unfortunately, high velocity resolution optical/UV observations 
are as yet not available for this system. We note, however, that the systemic 
velocity that we obtain for SBS~1543+593 is in excellent agreement with 
the value of $2868 \pm 2$~\kms~obtained from the single dish HI profile 
(\cite{bowen01b}). Further, the velocity obtained in our observations 
($2855\pm6$~\kms) for the gas coincident with the HII region (for which an optical 
emission spectrum exists) is in reasonable agreement with that obtained in the 
optical ($\sim$ 2700~\kms; \cite{reimers98}), given the poor resolution 
($18$~angstroms, i.e. $\sim 1000$~\kms) of the optical spectrum. Similarly, the 
velocity we obtain at the QSO location is $\sim 2870 \pm 10$~\kms, which 
agrees within the error bars with the velocity of $2700$~\kms~obtained from 
the Ly$-\alpha$ line (\cite{bowen01a}), given the large uncertainty 
($\sim 200-300$~\kms) in the latter measurement (\cite{bowen01b}). A high 
resolution low ionization metal line absorption spectrum of the QSO (and of Ton~1480, 
which lies behind NGC~4203) would provide interesting spot checks of the 
reliability in using such absorption lines to probe large scale gas kinematics 
of the absorbing galaxies (see, for example, \cite{prochaska97, prochaska98}).

	It is also interesting to compare the HI column density as obtained 
from the Ly-$\alpha$ spectrum with that obtained from the 21cm emission. The 
comparison is, however, complicated by the comparatively large size of the GMRT 
synthesized beam. At a distance of 38~Mpc,  $29^{''}$ corresponds to a linear 
size of $\sim 5.3$~kpc. Given this great disparity in the transverse sizes covered 
by the HI and UV measurements, the column density of $5 \times 10^{20}$ 
atoms~cm$^{-2}$ obtained from the 21cm map is in reasonable agreement with the 
estimate of $2.2\times10^{20}$ atoms~cm$^{-2}$ obtained from the Lyman-$\alpha$ profile.

	Comparisons of DLA samples with samples of galaxies selected by blind 
HI emission are complicated by issues of optical extinction. It has long 
been suggested that DLA samples could be seriously biased against high HI column 
density systems, because such absorbers would contain enough dust to substantially 
dim the background QSO (e.g. \cite{heisler88,fall93,pei99}). Deep optical
observations of a radio selected sample of QSOs however suggest that such biases 
may be modest (\cite{ellison01}). It is curious (but, of course, not statistically 
significant, particularly given the fact that optical depth effects may be less 
important in LSB galaxies than in normal spirals) that the QSO HS~1543+5921 lies behind a 
region of low local HI column density in SBS~1543+593. Interestingly, an inspection 
of the HI maps of the S0 galaxy NGC~4203 presented by van Driel et al. (1988) 
shows that Ton~1480 also lies behind a region of low column density. Unfortunately,
testing the above bias would require 21cm mapping of a statistically significant 
number of DLAs which is, alas, a task for the next generation radio telescope.
	
\begin{acknowledgement}
The GMRT observations presented in this paper would not have been possible 
without the many years of dedicated effort put in by the GMRT staff to build 
the telescope. The GMRT is run by the National Centre for Radio Astrophysics 
of the Tata Institute of Fundamental Research. Useful discussions with
R. Nityananda are gratefully acknowledged. We also thank the referee, 
E. Brinks, for several comments which improved both the readability and the
content of this paper.
\end{acknowledgement}

\end{document}